\documentclass[12pt]{iopart}

 \usepackage{iopams}
  \usepackage{amsthm}
\newfont{\myeu}{eurm10 at 12 pt}
\newfont{\bfrak}{eufb10 at 12 pt}
\newfont{\mycal}{eufb10 at 12pt}

\def\sfactor#1#2{\Bigg[\begin{array}{@{}c@{}}#1\\#2\end{array}\Bigg]}

\def\i{{\rm i}}

\def\be{\begin{equation}}
\def\ee{\end{equation}}
\def\ba{\begin{eqnarray}}
\def\ea{\end{eqnarray}}
\def\Btau2{{\boldsymbol \tau}^{\vphantom{'}}_{\!2}}

\def\sfactor#1#2{\Bigg[\begin{array}{@{}c@{}}#1\\#2\end{array}\Bigg]}

\def\sfactor#1#2{\Bigg[\begin{array}{@{}c@{}}#1\\#2\end{array}\Bigg]}
\def\bino#1#2{\Bigg(\begin{array}{@{}c@{}}#1\\#2\end{array}\Bigg)}

\def\halfs{{\scriptstyle{\frac 12}}}
\begin{document}

\title [Serre Relations in the Superintegrable Model]%
{Serre Relations in the Superintegrable Model}
\author{Helen Au-Yang and Jacques H H Perk%
}
\address{Department of Physics, Oklahoma State University,\\
145 Physical Sciences, Stillwater, OK 74078-3072, USA.}
\ead{\mailto{perk@okstate.edu}, \mailto{helenperk@yahoo.com}}

\begin{abstract}
We derive the Serre relations for the generators of the
quantum loop algebra $L({\mathfrak{sl}}_2)$
of the superintegrable $\Btau2$ model in $Q\ne0$ sectors,
thus proving a fundamental conjecture in an earlier paper
on the superintegrable chiral Potts model.
\end{abstract}

\vspace{2pc}

\section{Introduction}

In 1985 von Gehlen and Rittenberg \cite{vGR} introduced a special
hermitian quantum spin chain with $N$ states per site, having Ising-like
features and generalizing a 3-state model of Howes, Kadanoff and
den Nijs \cite{HKdN}. This model was later called superintegrable
\cite{AMPT89}, as the two terms in the hamiltonian generate an
Onsager algebra \cite{JHHP89} \textit{and} the Boltzmann weights of the
corresponding classical two-dimensional chiral Potts model satisfy
star-triangle (Yang--Baxter) relations \cite{AMPTY,BPAuY88}.

A quantum-group theoretical interpretation of the model was first
given by Bazhanov and Stroganov \cite{BS} when they introduced
the $\Btau2$ model connecting the integrable chiral Potts model
with the six-vertex model. More precisely, a square of four
chiral Potts Boltzmann weights \cite{BPAuY88} is the intertwiner
of cyclic representations of the affine quantum group
$U_q(\widehat{{\mathfrak{sl}}_2})$ \cite{JimboNK}, whereas the
six-vertex ${\cal R}$-matrix intertwines spin-$\frac12$
heighest-weight representations and the $\Btau2$ model weights
intertwine spin-$\frac12$ and cyclic representations. All this
is expressed in a sequence of Yang--Baxter equations involving
the intertwiners \cite{BS,BBP,Baxter-tau}.

Not only do superintegrable chiral Potts models  have Ising-like
spectra \cite{vGR,AMPT89,AMP,Baxsu}, for a periodic chain with
spin-shift quantum number $Q\!=\!0$ and chain length $L$
a multiple of $N$, it has been shown \cite{NiDe1,NiDe2} that the
eigenspace supports a quantum loop algebra $L({\mathfrak{sl}}_2)$.
Furthermore, this loop algebra can be decomposed into $r$ simple
${\mathfrak{sl}}_2$ algebras, with $r=m_0=(N-1)L/N$ for the
ground-state sector \cite{APsu1,APsu2,APsu4}\footnote{All equations
in~\cite{APsu2} are denoted here by prefacing IV
to its equation numbers.} 

We have also worked out the ground-state sector for $Q\!\ne\!0$ cases,
under the assumption that certain Serre relations hold \cite{APsu4}.
Even though we have shown that these relations hold when operated
on some special vectors (see Appendix B of \cite{APsu4}) and we
have also tested them extensively by computer for small systems,
a proof has been lacking up to now. In this paper, we shall present
the missing proof.

In section 1, we relate the operators used in our paper \cite {APsu4}
to generators of $U_q(\widehat{{\mathfrak{sl}}_2})$ and operators of
the quantum loop algebra $L({\mathfrak{sl}}_2)$ \cite{NiDe1,NiDe2}.
We then use the higher-order quantum Serre relations of Lusztig
\cite{Lusztig} to derive certain relations in section 2. Next we
rewrite these relations in terms of our operators in sections 3
and 4. We can then in section 5 use these relations to prove the
Serre relations for the generators used in \cite{APsu4}. We end
with a brief conclusion in section 6.
 
\section{Relationship between the generators}

In our paper \cite{APsu4}, the generators $\mbox{\mycal e}_j$ and
$\mbox{\mycal f}_j$ are defined in (IV.25), with $\bf Z$ and $\bf X$
given in (IV.20). These are different from the usual
$\mbox{\mycal e}'_j$ and $\mbox{\mycal f}'_j$ of the quantum group
$U_q(\widehat{{\mathfrak{sl}}_2})$ \cite{JimboNK},
but are related by \cite{NiDe1}
\be
\mbox{\mycal e}'_j=-q\mbox{\mycal e}_j{\bf Z}_j^{-1/2},\quad
\mbox{\mycal f}'_j=q{\bf Z}_j^{-1/2}\mbox{\mycal f}_j,\quad 
\mbox{\mycal k}'_j=q^{-1}{\bf Z}_j^{-1},\quad
\omega=q^2=\e^{2\pi\i/N}.
\label{efk}\ee
Substituting these into the operators ${\bf B}_{\pm}$ and 
${\bf C}_{\pm}$ defined on page 368 of \cite{NiDe1}, and comparing
with $\bar{\bf B}_{1}$, $\bar{\bf B}_{L}$, $\bar{\bf C}_{0}$ and
$\bar{\bf C}_{L-1}$ defined in (IV.24) and (IV.55), we find
\ba
&{\bf C}_{+}=-q^{-L+2}\bar{\bf C}_{0}{\bf A}^{-\halfs}_L,\quad
&{\bf B}_{+}=q^{-L+2}{\bf A}^{-\halfs}_L\bar{\bf B}_{1}, 
\quad {\bf A}_L=\prod_{i=1}^L{\bf Z}_i,\cr
&{\bf C}_{-}=-q\bar{\bf C}_{L-1}{\bf A}^{-\halfs}_L,\quad
&{\bf B}_{-}=q{\bf A}^{-\halfs}_L\bar{\bf B}_{L}.
\label{CB}\ea
From now on, we drop the bars from the $\bf B$ and $\bf C$ symbols
taken from \cite{APsu4}. Defining
\ba
{\bf C}^{(n)}_{\pm}=\frac{{\bf C}^n_{\pm}}{[n]_q!},\quad
&{\bf B}^{(n)}_{\pm}=\frac{{\bf B}^n_{\pm}}{[n]_q!},\quad
&[n]_q!=\prod_{i=1}^n\frac{q^i-q^{-i}}{q-q^{-1}},\cr
{\bf C}^{(n)}_{\ell}=\frac{{\bf C}^n_{\ell}}{[n]!},\quad
&{\bf B}^{(n)}_{\ell}=\frac{{\bf B}^n_{\ell}}{[n]!},\quad
&[n]!=\prod_{i=1}^n\frac{1-\omega^{i}}{1-\omega},\quad
\ell=0,1,L-1,L,
\label{defCB}\ea
we find from (\ref{CB}) and (\ref{defCB}) the relations
\ba
\fl [n]_q!=q^{-\halfs n(n-1)}[n]!,\quad
{\bf C}^{(n)}_{-}=(-1)^n{\bf A}^{-\halfs n}_L{\bf C}^{(n)}_{L-1},\quad
{\bf B}^{(n)}_{-}={\bf B}^{(n)}_{L}{\bf A}^{-\halfs n}_L,\cr
\quad\,{\bf C}^{(n)}_{+}=
(-1)^n q^{n(1-L)}{\bf A}^{-\halfs n}_L{\bf C}^{(n)}_{0},\quad
{\bf B}^{(n)}_{+}=q^{n(1-L)}{\bf B}^{(n)}_{1}{\bf A}^{-\halfs n}_L.
\label{ClBn}\ea

\section{Higher-order Serre relations}

We follow the conventions of Nishino and Deguchi \cite {NiDe1} letting
\be
{\bf E}_0={\bf B}_+,\quad {\bf E}_1={\bf C}_+,\quad
{\bf F}_0={\bf C}_-,\quad {\bf F}_1={\bf B}_-,
\ee
so that we may adapt Chapter 7 of Lusztig \cite{Lusztig} and define
the following function for the cyclic case with $q^{2N}=1$,
\be\fl
f_{i,j,n,m}=f_{n,m}=\sum_{r+s=m}^m (-1)^r q^{r(2n-m+1)}\theta_i^{(r)}
\theta_j^{(n)}\theta_i^{(s)},\quad i,j=0,1,\quad j\ne i,
\label{fij}\ee
where we may choose $\theta_i={\bf E}_i$ or $\theta_i={\bf F}_i$.
It is shown by Lusztig in Proposition 7.15.(b) \cite{Lusztig} that
if $m>2n$, then $f_{n,m}=0$. For $n=1$ and $m=3$, these are the usual
quantum Serre relations given in (3.23) through (3.26) of \cite{DFM}.%
\footnote{Use translation $S^-=B^-$, $T^-=B^+$, $S^+=C^+$, $T^+=C^-$.}

We follow the steps of Lusztig's proof. Let us first consider the case
$m-2n\ge N$, so that $f_{n,m-\ell}=0$ for $\ell\le N-1\le m-2n-1$.
Consequently we have
\be
g=\sum_{\ell=0}^{N-1}(-1)^\ell q^{\ell(1-m)}
f_{n,m-\ell}\,\theta_i^{(\ell)}=0.
\ee
Using (\ref{fij}), we find
\ba
g&=\sum_{\ell=0}^{N-1}\sum_{r+s'=m-\ell}(-1)^{\ell+r}
q^{\ell(1-m)+r(2n-m+\ell+1)}
\theta_i^{(r)}\theta_j^{(n)}\theta_i^{(s')}\theta_i^{(\ell)}\cr
&=\sum_{s=0}^m c_{s}\,
\theta_i^{(m-s)}\theta_j^{(n)}\theta_i^{(s)}=0,\qquad r=m-s,
\label{gh}\ea
where
\ba
\fl\qquad c_{s}=\sum_{\ell=0}^{N-1}(-1)^{\ell+m-s}
q^{\ell(1-s)+(m-s)(2n-m+1)}
\sfactor s{\ell}_q,\quad \sfactor s{\ell}_q=
\frac{[s]_q!}{[\ell]_q![s-\ell]_q!}.
\label{cs}\ea
These are exactly the same as in Lusztig. But from now on, we will
use the cyclic property as in \cite{DFM}.
We set $s=kN+p$ for $0\le k\le \lfloor m/N\rfloor$,
with $0\le p\le N-1$ if $0\le k\le \lfloor m/N\rfloor-1$,
and $0\le p\le m-N \lfloor m/N\rfloor$ if $k=\lfloor m/N\rfloor$.
Using (3.55) of \cite{DFM}, namely
\be 
\sfactor s\ell_q=\sfactor{kN+p}\ell_q=
q^{kN\ell}\sfactor p\ell_q
\label{kNp}\ee
we rewrite $c_s$ in (\ref{cs}) as
\ba
c_{kN+p}&=(-1)^{m-kN-p}q^{(m-kN-p)(2n-m+1)}
\sum_{\ell=0}^{p}(-1)^{\ell} q^{\ell(1-p)}
\sfactor p{\ell}_q\cr
&=(-1)^{m-kN-p}q^{(m-kN-p)(2n-m+1)}\delta_{p,0},
\label{cjNp}\ea
where 1.3.4 of Lusztig \cite{Lusztig}, or (3.58) of \cite{DFM} is used. Substituting this equation into (\ref{gh}), we find
\be\fl
(-1)^{m}q^{m(2n-m+1)} \Bigg[\theta_i^{(m)}\theta_j^{(n)}+
\sum_{k=1}^{\lfloor m/N\rfloor} (-1)^{k(N+m-1)}
\theta_i^{(m-kN)}\theta_j^{(n)}\theta_i^{(kN)}\Bigg]=0.
\label{id1}\ee
Particularly, letting $\theta_i={\bf B}_\pm$ and
$\theta_j={\bf C}_\pm$ and $n=Q$, $m=2N+Q$, so that
$m-2n=2N-Q>N$, we find the identity
\be
{\bf B}_\pm^{(2N+Q)}{\bf C}_\pm^{(Q)}+(-1)^{(N+Q-1)}
{\bf B}_\pm^{(N+Q)}{\bf C}_\pm^{(Q)}{\bf B}_\pm^{(N)}+
{\bf B}_\pm^{(Q)}{\bf C}_\pm^{(Q)}{\bf B}_\pm^{(2N)}=0.
\label{BCN}\ee
Interchanging $i$ and $j$, we have
\be
{\bf C}_\pm^{(2N+Q)}{\bf B}_\pm^{(Q)}+(-1)^{(N+Q-1)}
{\bf C}_\pm^{(N+Q)}{\bf B}_\pm^{(Q)}{\bf C}_\pm^{(N)}+
{\bf C}_\pm^{(Q)}{\bf B}_\pm^{(Q)}{\bf C}_\pm^{(2N)}=0.
\label{CBN}\ee

Next we consider the case that $0\le m-2n\le N-1$. Let
\ba
g=\sum_{\ell=0}^{m-2n-1}(-1)^\ell q^{\ell(1-m)}\,f_{n,m-\ell}
\theta_i^{(\ell)}
=\sum_{s=0}^m c_{s}\,
\theta_i^{(m-s)}\theta_j^{(n)}\theta_i^{(s)}=0,
\label{ghh}\ea
where
\ba
c_{s}=(-1)^{m-s}q^{(m-s)(2n-m+1)}
\sum_{\ell=0}^{m-2n-1}(-1)^{\ell} q^{\ell(1-s)}\sfactor s{\ell}_q.
\label{csn}\ea
Now if we again write $s=kN+p$, then for $0\le p\le m-2n-1$,
$c_s$ is again summable and is given by (\ref{cjNp}).
However, for $m-2n\le p\le N-1$, the sum in (\ref{csn}) is
not summable. Nevertheless, since
\be
\theta_i^{(kN+p)}\theta_i^{(N-m+2n)}=
\sfactor{kN+N+p-m+2n}{N-m+2n}_q\theta_i^{(kN+N+p-m+2n)},
\ee
and for $m-2n\le p\le N-1$, we have
\be
\sfactor{kN+N+p-m+2n}{N-m+2n}_q=
q^{(N-m+2n)N(k+1)}\sfactor{p-m+2n}{N-m+2n}_q=0
\ee
we find
\be
\sum_{p=m-2n}^{N-1}c_{kN+p}\,
\theta_i^{(m-kN-p)}\theta_j^{(n)}\theta_i^{(kN+p)}
\theta_i^{(N-m+2n)}=0.
\ee
Thus by multiplying $\theta_i^{(N-m+2n)}$ to $g$ on the right,
we may get rid of the terms involving these unsummable $c_s$,
and find
\ba
0=g\,\theta_i^{(N-m+2n)}=\sum_{k=0}^{\lfloor m/N\rfloor}c_{kN}
\theta_i^{(m-kN)}\theta_j^{(n)}
\theta_i^{(kN)}\theta_i^{(N-m+2n)}\cr
=\sum_{k=0}^{\lfloor m/N\rfloor}c_{kN}
\theta_i^{(m-kN)}\theta_j^{(n)}
\theta_i^{(kN+N-m+2n)}\sfactor{kN+N-m+2n}{N-m+2n}_q\cr
=(-1)^{m}q^{m(2n-m+1)} \Bigg\{\sum_{k=0}^{\lfloor m/N\rfloor}
(-1)^{k}\theta_i^{(m-kN)}\theta_j^{(n)}
\theta_i^{(kN+N-m+2n)}\Bigg\}.
\label{id2}\ea

If we let $n=Q$ and $m=N+Q$, so that $m-2n=N-Q>0$,
then (\ref{id2}) becomes
\ba\fl
{\bf B}_\pm^{(N+Q)}{\bf C}_\pm^{(Q)}{\bf B}_\pm^{(Q)}=
{\bf B}_\pm^{(Q)}{\bf C}_\pm^{(Q)}{\bf B}_\pm^{(N+Q)},\qquad
{\bf C}_\pm^{(N+Q)}{\bf B}_\pm^{(Q)}{\bf C}_\pm^{(Q)}=
{\bf C}_\pm^{(Q)}{\bf B}_\pm^{(Q)}{\bf C}_\pm^{(N+Q)}.
\label{BCB}\ea

Now let $n=N+Q$ and $m=3N+Q$. Again we have $m-2n=N-Q>0$,
and for such values (\ref{id2})
becomes
\ba\fl
{\bf B}_\pm^{(3N+Q)}{\bf C}_\pm^{(N+Q)}{\bf B}_\pm^{(Q)}&-
{\bf B}_\pm^{(2N+Q)}{\bf C}_\pm^{(N+Q)}{\bf B}_\pm^{(N+Q)}\cr
&+{\bf B}_\pm^{(N+Q)} {\bf C}_\pm^{(N+Q)}{\bf B}_\pm^{(2N+Q)}-
{\bf B}_\pm^{(Q)}{\bf C}_\pm^{(N+Q)}{\bf B}_\pm^{(Q+3N)}=0,
\label{BCBC}\ea
or
\ba\fl
{\bf C}_\pm^{(3N+Q)}{\bf B}_\pm^{(N+Q)}{\bf C}_\pm^{(Q)}&-
{\bf C}_\pm^{(2N+Q)}{\bf B}_\pm^{(N+Q)}{\bf C}_\pm^{(N+Q)}\cr
&+{\bf C}_\pm^{(N+Q)} {\bf B}_\pm^{(N+Q)}{\bf C}_\pm^{(2N+Q)}-
{\bf C}_\pm^{(Q)}{\bf B}_\pm^{(N+Q)}{\bf C}_\pm^{(Q+3N)}=0.
\label{CBCB}\ea
If $Q=0$, these are the Serre relations given by
(3.31) through (3.34) in \cite{DFM}.

\section{Alternative form}

Substituting (\ref{ClBn}) into (\ref{BCN}) and (\ref{CBN}), and using
the commutation relations
\be
A_L^{-\halfs}{\bf C}_\ell=q{\bf C}_\ell A_L^{-\halfs},\quad
{\bf B}_\ell A_L^{-\halfs}=qA_L^{-\halfs}{\bf B}_n,
\quad\ell=0, L-1,\quad n=1,L,
\label{Commu}\ee
we find
\ba
{\bf B}_1^{(2N+Q)}{\bf C}_0^{(Q)}-
{\bf B}_1^{(N+Q)}{\bf C}_0^{(Q)}{\bf B}_1^{(N)}+
{\bf B}_1^{(Q)}{\bf C}_0^{(Q)}{\bf B}_1^{(2N)}&=0,
\label{B1C0N}\\
{\bf C}_0^{(2N+Q)}{\bf B}_1^{(Q)}-
{\bf C}_0^{(N+Q)}{\bf B}_1^{(Q)}{\bf C}_0^{(N)}+
{\bf C}_0^{(Q)}{\bf B}_1^{(Q)}{\bf C}_0^{(2N)}&=0.
\label{C0B1N}\ea
Similarly, substituting (\ref{ClBn}) into (\ref{BCB}) and using
(\ref{Commu}), we obtain
\ba\fl
{\bf B}_1^{(N+Q)}{\bf C}_0^{(Q)}{\bf B}_1^{(Q)}=
{\bf B}_1^{(Q)}{\bf C}_0^{(Q)}{\bf B}_1^{(N+Q)},\qquad
{\bf C}_0^{(N+Q)}{\bf B}_1^{(Q)}{\bf C}_0^{(Q)}=
{\bf C}_0^{(Q)}{\bf B}_1^{(Q)}{\bf C}_0^{(N+Q)}.
\label{B1C0B1}\ea
Finally, from (\ref{BCBC}), (\ref{CBCB}) together with
(\ref{ClBn}) and (\ref{Commu}), we get
\ba\fl
{\bf B}_1^{(3N+Q)}{\bf C}_0^{(N+Q)}{\bf B}_1^{(Q)}&-
{\bf B}_1^{(2N+Q)}{\bf C}_0^{(N+Q)}{\bf B}_1^{(N+Q)}\cr
&+{\bf B}_1^{(N+Q)} {\bf C}_0^{(N+Q)}{\bf B}_1^{(2N+Q)}-
{\bf B}_1^{(Q)}{\bf C}_0^{(N+Q)}{\bf B}_1^{(Q+3N)}=0.
\label{B1C0BC}\\
\fl
{\bf C}_0^{(3N+Q)}{\bf B}_1^{(N+Q)}{\bf C}_0^{(Q)}&-
{\bf C}_0^{(2N+Q)}{\bf B}_1^{(N+Q)}{\bf C}_0^{(N+Q)}\cr
&+{\bf C}_0^{(N+Q)} {\bf B}_1^{(N+Q)}{\bf C}_0^{(2N+Q)}-
{\bf C}_0^{(Q)}{\bf B}_1^{(N+Q)}{\bf C}_0^{(Q+3N)}=0.
\label{C0B1CB}\ea
Similar equations hold if we replace ${\bf B}_1$ by ${\bf B}_L$
and ${\bf C}_0$ by ${\bf C}_{L-1}$.

\section{Serre relations for the generators of the loop algebra}

We will now prove the Serre relations (IV.90) for the generators
given in (IV.88), i.e.,
\be
{\bf x}^{-}_{1,Q}=
{\bf C}_0^{(Q)}{\bf B}_1^{( N+Q)},\quad
{\bf x}^{+}_{0,Q}=
{\bf C}_0^{(N+Q)}{\bf B}_1^{(Q)},
\label{xmpo}
\ee
where we have dropped the common constant factors for convenience.
We use first the equation on the right and then the one on the left
in (\ref{B1C0B1}) to find 
\ba
{\bf x}^{+}_{0,Q}{\bf x}^{-}_{1,Q}&=
[{\bf C}_0^{(N+Q)}{\bf B}_1^{(Q)}{\bf C}_0^{(Q)}]{\bf B}_1^{( N+Q)}\cr
&={\bf C}_0^{(Q)}{\bf B}_1^{(Q)}{\bf C}_0^{(N+Q)}{\bf B}_1^{( N+Q)}=
{\bf C}_0^{(N+Q)}{\bf B}_1^{(N+Q)}{\bf C}_0^{(Q)}{\bf B}_1^{(Q)}.
\label{xpm}\ea
This means
\be
[{\bf C}_0^{(Q)}{\bf B}_1^{(Q)},{\bf C}_0^{(N+Q)}{\bf B}_1^{( N+Q)}]=0.
\label{commu}\ee
It is easy to verify that
\ba\fl
{\bf B}_1^{( kN+Q)}{\bf B}_1^{( jN)}=
\sfactor{kN+jN+Q}{kN+Q}{\bf B}_1^{( jN+kN+Q)}
=\bino{k+j}{k}{\bf B}_1^{( jN+kN+Q)}.
\label{mulo}\ea
We again use (\ref{B1C0B1}) and (\ref{mulo}) to find
\ba
({\bf x}^{-}_{1,Q})^2&=
{\bf C}_0^{(Q)}[{\bf B}_1^{(N+Q)}{\bf C}_0^{(Q)}{\bf B}_1^{( Q)}]
{\bf B}_1^{( N)}=
2{\bf C}_0^{(Q)}{\bf B}_1^{(Q)}{\bf C}_0^{(Q)}{\bf B}_1^{(2 N+Q)}
\label{xm2a}\\
&={\bf C}_0^{(Q)}{\bf B}_1^{( N)}
[{\bf B}_1^{(Q)}{\bf C}_0^{(Q)}{\bf B}_1^{( N+Q)}]
=2{\bf C}_0^{(Q)}{\bf B}_1^{(2N+Q)}{\bf C}_0^{(Q)}{\bf B}_1^{( Q)}.
\label{xm2b}\ea
As a consequence, we obtain another identity,
\ba
{\bf C}_0^{(Q)}{\bf B}_1^{(Q)}{\bf C}_0^{(Q)}{\bf B}_1^{(2 N+Q)}=
{\bf C}_0^{(Q)}{\bf B}_1^{( 2N+Q)}{\bf C}_0^{(Q)}{\bf B}_1^{(Q)}.
\label{C0B1N2}\ea
Multiplying (\ref{xpm}) and (\ref{xm2a}) we obtain
\ba
{\bf x}^{+}_{0,Q}({\bf x}^{-}_{1,Q})^3=
2{\bf C}_0^{(Q)}{\bf B}_1^{(Q)}{\bf C}_0^{(N+Q)}{\bf B}_1^{(N+Q)}
{\bf C}_0^{(Q)}{\bf B}_1^{(Q)}{\bf C}_0^{(Q)}
{\bf B}_1^{(Q)}{\bf B}_1^{(2N)}.
\ea
Using (\ref{commu}) repeatedly to move operators with higher
exponents to the right, and then using (\ref{mulo}), we find
\ba
{\bf x}^{+}_{0,Q}({\bf x}^{-}_{1,Q})^3=
6{\bf C}_0^{(Q)}{\bf B}_1^{(Q)}{\bf C}_0^{(Q)}{\bf B}_1^{(Q)}
{\bf C}_0^{(Q)}{\bf B}_1^{(Q)}{\bf C}_0^{(N+Q)}{\bf B}_1^{(3N+Q)}.
\label{xpxm3}\ea
Similarly, by repeatedly using (\ref{B1C0B1}), and then
(\ref{commu}), we also find
\ba\fl
({\bf x}^{-}_{1,Q}){\bf x}^{+}_{0,Q}({\bf x}^{-}_{1,Q})^2&=
{\bf C}_0^{(Q)}[{\bf B}_1^{(N+Q)}{\bf C}_0^{(Q)}{\bf B}_1^{(Q)}]
{\bf C}_0^{(N+Q)}[{\bf B}_1^{(N+Q)}{\bf C}_0^{(Q)}{\bf B}_1^{(Q)}]
{\bf B}_1^{(N)}\cr
&=2{\bf C}_0^{(Q)}{\bf B}_1^{(Q)}{\bf C}_0^{(Q)}{\bf B}_1^{(N+Q)}
{\bf C}_0^{(N+Q)}{\bf B}_1^{(Q)}{\bf C}_0^{(Q)}{\bf B}_1^{(2N+Q)}\cr
&=2{\bf C}_0^{(Q)}{\bf B}_1^{(Q)}{\bf C}_0^{(Q)}{\bf B}_1^{(Q)}
{\bf C}_0^{(Q)}{\bf B}_1^{(N+Q)}{\bf C}_0^{(N+Q)}{\bf B}_1^{(2N+Q)}.
\label{xmxpxm2}\ea
From (\ref{xpm}), (\ref{xm2a}) and (\ref{C0B1N2}), we obtain
\ba\fl
({\bf x}^{-}_{1,Q})^2({\bf x}^{+}_{0,Q}{\bf x}^{-}_{1,Q})&=
2{\bf C}_0^{(Q)}{\bf B}_1^{(Q)}[{\bf C}_0^{(Q)}{\bf B}_1^{(2N+Q)}
{\bf C}_0^{(Q)}{\bf B}_1^{(Q)}]{\bf C}_0^{(N+Q)}{\bf B}_1^{(N+Q)}\cr
&=2{\bf C}_0^{(Q)}{\bf B}_1^{(Q)}{\bf C}_0^{(Q)}{\bf B}_1^{(Q)}
{\bf C}_0^{(Q)}{\bf B}_1^{(2N+Q)}{\bf C}_0^{(N+Q)}{\bf B}_1^{(N+Q)}.
\label{xm2xpxm}\ea
Now we use (\ref{C0B1N2}) and (\ref{mulo}) to get
\ba\fl\qquad
({\bf x}^{-}_{1,Q})^3{\bf x}^{+}_{0,Q}&=
2{\bf C}_0^{(Q)}{\bf B}_1^{(Q)}[{\bf C}_0^{(Q)}{\bf B}_1^{(2N+Q)}
{\bf C}_0^{(Q)}{\bf B}_1^{(Q)}]{\bf B}_1^{(N)}
{\bf C}_0^{(N+Q)}{\bf B}_1^{(Q)}\cr
&=6{\bf C}_0^{(Q)}{\bf B}_1^{(Q)}{\bf C}_0^{(Q)}{\bf B}_1^{(Q)}
{\bf C}_0^{(Q)}{\bf B}_1^{(3N+Q)}{\bf C}_0^{(N+Q)}{\bf B}_1^{Q)}.
\label{xm3xp}\ea
Finally, combining all these, we find
\ba\fl
[[[{\bf x}^{+}_{0,Q},{\bf x}^{-}_{1,Q}],{\bf x}^{-}_{1,Q}],
{\bf x}^{-}_{1,Q}]\cr\fl
={\bf x}^{+}_{0,Q}({\bf x}^{-}_{1,Q})^3-
3({\bf x}^{-}_{1,Q}){\bf x}^{+}_{0,Q}({\bf x}^{-}_{1,Q})^2
+3({\bf x}^{-}_{1,Q})^2({\bf x}^{+}_{0,Q}{\bf x}^{-}_{1,Q})-
({\bf x}^{-}_{1,Q})^3{\bf x}^{+}_{0,Q}\cr\fl
=6{\bf C}_0^{(Q)}{\bf B}_1^{(Q)}{\bf C}_0^{(Q)}{\bf B}_1^{(Q)}
{\bf C}_0^{(Q)}[{\bf B}_1^{(Q)}{\bf C}_0^{(N+Q)}{\bf B}_1^{(3N+Q)}-
{\bf B}_1^{(N+Q)}{\bf C}_0^{(N+Q)}{\bf B}_1^{(2N+Q)}\cr
\hspace{0.87in}
-{\bf B}_1^{(2N+Q)}{\bf C}_0^{(N+Q)}{\bf B}_1^{(N+Q)}+
{\bf B}_1^{(3N+Q)}{\bf C}_0^{(N+Q)}{\bf B}_1^{Q)}]=0,
\label{serrea}\ea
as seen from (\ref{B1C0BC}).

It is straightforward to show that
\ba
{\bf x}^{-}_{1,Q}({\bf x}^{+}_{0,Q})^3&=
6{\bf C}_0^{(Q)}{\bf B}_1^{(N+Q)}{\bf C}_0^{(3N+Q)}{\bf B}_1^{(Q)}
{\bf C}_0^{(Q)}{\bf B}_1^{(Q)}{\bf C}_0^{(Q)}{\bf B}_1^{Q)}.
\label{xmxp3}\\
{\bf x}^{+}_{0,Q}{\bf x}^{-}_{1,Q}({\bf x}^{+}_{0,Q})^2&=
2{\bf C}_0^{(N+Q)}{\bf B}_1^{(N+Q)}{\bf C}_0^{(2N+Q)}
{\bf B}_1^{(Q)}{\bf C}_0^{(Q)}{\bf B}_1^{(Q)}
{\bf C}_0^{(Q)}{\bf B}_1^{Q)}.
\label{xpxmxp2}\\
({\bf x}^{+}_{0,Q})^2{\bf x}^{-}_{1,Q}{\bf x}^{+}_{0,Q}&=
2{\bf C}_0^{(2N+Q)}{\bf B}_1^{(N+Q)}{\bf C}_0^{(N+Q)}{\bf B}_1^{(Q)}
{\bf C}_0^{(Q)}{\bf B}_1^{(Q)}{\bf C}_0^{(Q)}{\bf B}_1^{Q)}.
\label{xp2xmxp}\\
({\bf x}^{+}_{0,Q})^3{\bf x}^{-}_{1,Q}&=6{\bf C}_0^{(3N+Q)}
{\bf B}_1^{(N+Q)}{\bf C}_0^{(Q)}
{\bf B}_1^{(Q)}{\bf C}_0^{(Q)}{\bf B}_1^{(Q)}{\bf C}_0^{(Q)}
{\bf B}_1^{Q)}.
\label{xp3xm}
\ea
Consequently, we use (\ref{C0B1CB}) to show that
\be
[[[{\bf x}^{-}_{1,Q},{\bf x}^{+}_{0,Q}],{\bf x}^{+}_{0,Q}],
{\bf x}^{+}_{0,Q}]=0.
\label{serreb}\ee

We have also defined \cite{APsu6} the generators
\be
{\bf\bar x}^{-}_{0,Q}=
{\bf B}_L^{( N+Q)}{\bf C}_{L-1}^{(Q)},\quad
{\bf \bar x}^{+}_{-1,Q}=
{\bf B}_L^{( Q)}{\bf C}_{L-1}^{(N+Q)}.
\label{bxmpo}\ee
Since (\ref{B1C0B1}), (\ref{B1C0BC}) and (\ref{C0B1CB}) also
hold if we replace ${\bf B}_1$ by ${\bf B}_L$ and
${\bf C}_0$ by ${\bf C}_{L-1}$, we can follow the same steps to prove
\ba
[[[{\bf\bar x}^{-}_{0,Q},{\bf \bar x}^{+}_{-1,Q}],
{\bf\bar x}^{+}_{-1,Q}],{\bf \bar x}^{+}_{-1,Q}]=0,\quad
[[[{\bf \bar x}^{+}_{-1,Q},{\bf\bar x}^{-}_{0,Q}],
{\bf \bar x}^{-}_{0,Q}],{\bf\bar x}^{-}_{0,Q}]=0.
\label{serrec}\ea

\section{Conclusion}

The two Serre relations (IV.90) conjectured in \cite{APsu4} have now
been proved, see (\ref{serrea}) and (\ref{serreb}). Two other Serre
relations (\ref{serrec}) applicable to the quantum subalgebra related
to the state $|\bar\Omega\rangle$ \cite{APsu4}, rather than
$|\Omega\rangle$, have also been derived. The Serre relations (32)
in \cite{APsu1} are included as the special case $Q=0$,
for which the two subalgebras combine to one quantum loop algebra.

\section*{Acknowledgments}

This work was supported in part by the National Science Foundation
under grant No.\ PHY-07-58139.


\section*{References}
\begin{thebibliography}{000}
%
\bibitem{vGR}{%
von Gehlen G and Rittenberg V 1985
\textrm{$Z_n$-symmetric quantum chains with an infinite set of
conserved charges and $Z_n$ zero modes}
\textit{Nucl. Phys. B} {\bf 257} 351--70}
%
\bibitem{HKdN}{%
Howes S, Kadanoff L P and Den Nijs M 1983
\textrm{Quantum model for commen\-surate-incommen\-surate transitions}
\textit{Nucl. Phys. B} {\bf 215} [FS7] 169--208}
%
\bibitem{AMPT89}{%
Albertini G, McCoy B M, Perk J H H and Tang S 1989
\textrm{Excitation spectrum and order parameter for the integrable
$N$-state chiral Potts model}
\textit{Nucl. Phys. B} {\bf 314} 741--63}
%
\bibitem{JHHP89}{%
Perk J H H 1989
\textrm{Star-triangle equations, quantum Lax pairs,
and higher genus curves}
\textit{Proc. 1987 Summer Research Institute on Theta Functions}
[\textit{Proc. Symp. Pure Math.} {\bf 49} part 1]
(Providence, RI: Am. Math. Soc.) pp 341--54}
%
\bibitem{AMPTY}{%
Au-Yang H, McCoy B M, Perk J H H, Tang S and Yan M-L 1987
\textrm{Commuting transfer matrices in the chiral Potts models:
Solutions of the star-triangle equations with genus $> 1$}
\textit{Phys. Lett.} A {\bf 123} 219--23}
%
\bibitem{BPAuY88}{%
Baxter R J, Perk J H H and Au-Yang H 1988
\textrm{New solutions of the star-triangle relations for the chiral
Potts model}
\textit{Phys. Lett.} A {\bf 128} 138--42}
%
\bibitem{BS}{%
Bazhanov V V and Stroganov Yu G 1990
\textrm{Chiral Potts model as a descendent of the six-vertex model}
\textit{J. Stat. Phys.} {\bf 59} 799--817}
%
\bibitem{JimboNK}{%
Jimbo M 1992
\textrm{Topics from representations of $U_q(\mathfrak g)$---%
an introductory guide to physicists}
\textit{Quantum Groups and Quantum Integrable Systems
(Nankai Lectures on Mathematical Physics)}
ed M-L Ge (Singapore: World Scientific) pp~1--61}
%
\bibitem{BBP}{%
Baxter R J, Bazhanov V V and Perk J H H 1990
\textrm{Functional relations for transfer matrices of the
chiral Potts model}
\textit{Int. J. Mod. Phys.} B {\bf 4} 803--70}
%
\bibitem{Baxter-tau}{%
Baxter R J 2004
\textrm{Transfer matrix functional relations for the
generalized $\tau_2(t_q)$ model}
\textit{J. Stat. Phys.} {\bf 117} 1--25
(arXiv:cond-mat/0409493)}
%
\bibitem{AMP}{%
Albertini G, McCoy B M and Perk J H H 1989
\textrm{Eigenvalue spectrum of the superintegrable
chiral Potts model}
\textit{Adv. Stud. Pure Math.} vol 19
(Tokyo: Kinokuniya Academic) pp 1--55}
%
\bibitem{Baxsu}{%
Baxter R J 1989
\textrm{Superintegrable chiral Potts model: thermodynamic
properties, an ``inverse" model, and a simple associated Hamiltonian}
\textit{J. Stat. Phys.} {\bf 57} 1--39}
%
\bibitem{NiDe1}{%
Nishino A and Deguchi T 2006
\textrm{The $L({\mathfrak sl}_2)$ symmetry of the Bazhanov--Stroganov
model associated with the superintegrable chiral Potts model}
\textit{Phys. Lett.} A {\bf 356} 366--70
(arXiv:cond-mat/0605551)}
%
\bibitem{NiDe2}{%
Nishino A and Deguchi T 2008
\textrm{An algebraic derivation of the eigenspaces associated with an
Ising-like spectrum of the superintegrable chiral Potts model}
\textit{J. Stat. Phys.} {\bf 133} 587--615
(arXiv:0806.1268)}
%
\bibitem{APsu1}{%
Au-Yang H and Perk J H H 2008
\textrm{Eigenvectors in the superintegrable model I:
${\mathfrak{sl}}_2$ generators}
\textit{J. Phys. A: Math. Theor.} {\bf 41} 275201 (10pp)
(arXiv:0710.5257)}
%
\bibitem{APsu2}{%
Au-Yang H and Perk J H H 2009
\textrm{Eigenvectors in the superintegrable model II:
ground state sector}
\textit{J. Phys. A: Math. Theor.} {\bf 42} 375208 (16pp)
(arXiv:0803.3029)}
%
\bibitem{APsu4}{%
Au-Yang H and Perk J H H 2011
\textrm{Quantum loop subalgebra and eigenvectors of the
superintegrable chiral Potts transfer matrices}
\textit{J. Phys. A: Math. Theor.} {\bf 44} 025205 (26pp)
(arXiv:0907.0362)}
%
\bibitem{DFM}{%
Deguchi T, Fabricius K and McCoy B M 2001
\textrm{The $sl_2$ loop algebra symmetry of the six-vertex model at
roots of unity}
\textit{J. Stat. Phys.} {\bf 102} 701--36
(arXiv:cond-mat/9912141)}
%
\bibitem{Lusztig}{%
Lusztig G 1993
\textit{Introduction to Quantum Groups}
(Boston: Birkh\"auser) ch~7}
%
\bibitem{APsu6}{%
Au-Yang H and Perk J H H 2012
\textrm{Finite size calculation of eigenvalues in the
superintegrable $\Btau2$ model and the eigenvectors of the
superintegrable chiral Potts models}
\textit{Preprint} in preparation}

\end{thebibliography}
\end{document}